\documentclass[journal]{IEEEtran}
\ifCLASSINFOpdf
\else
\fi

\usepackage{booktabs}

\usepackage[utf8]{inputenc}
\usepackage{amsmath}
\usepackage{mathtools}
\usepackage{braket}
\usepackage{graphicx}
\graphicspath{{./figures/}}
\usepackage{amsfonts}
\usepackage{algorithm}
\usepackage{algpseudocode}
\let\oldnl\nl
\newcommand{\nonl}{\renewcommand{\nl}{\let\nl\oldnl}}

\usepackage{csquotes}
\usepackage{hhline}
\usepackage{amssymb}
\usepackage{listings}
\usepackage{color}
\definecolor{codegreen}{rgb}{0,0.6,0}
\definecolor{codegray}{rgb}{0.5,0.5,0.5}
\definecolor{codepurple}{rgb}{0.58,0,0.82}
\definecolor{backcolour}{rgb}{0.95,0.95,0.92}

\usepackage{subcaption}
\usepackage{titlesec}

\usepackage{dcolumn}
\usepackage{tabularx}
\usepackage{hyperref}
\hypersetup{
    colorlinks=true,
    linktoc=all,
    linkcolor=black,
    citecolor=magenta,
}
\usepackage{longtable}
\usepackage{braket}
\newcolumntype{C}{>{\centering\arraybackslash}X}
\usepackage[font=small]{caption}
\usepackage{tikz,xcolor,hyperref}
\definecolor{lime}{HTML}{A6CE39}
\DeclareRobustCommand{\orcidicon}{%
    \begin{tikzpicture}
    \draw[lime, fill=lime] (0,0) 
    circle [radius=0.16] 
    node[white] {{\fontfamily{qag}\selectfont \tiny ID}};    \draw[white, fill=white] (-0.0625,0.095) 
    circle [radius=0.007];    \end{tikzpicture}
    \hspace{-2mm}}
\foreach \x in {A, ..., Z}{%
    \expandafter\xdef\csname orcid\x\endcsname{\noexpand\href{https://orcid.org/\csname orcidauthor\x\endcsname}{\noexpand\orcidicon}}
    }

\begin{document}
\hyphenation{op-tical net-works semi-conduc-tor}

\title{QSVM-QNN: Quantum Support Vector Machine Based Quantum Neural Network Learning Algorithm for Brain-Computer Interfacing Systems}
\author{Bikash~K.~Behera\thanks{B.~K. Behera is with the Bikash's Quantum (OPC) Pvt. Ltd., Mohanpur, WB, 741246 India, e-mail: (bikas.riki@gmail.com).}, Saif Al-Kuwari\thanks{Saif Al-Kuwari is with the Qatar Center for Quantum Computing, College of Science and Engineering, Hamad Bin Khalifa University, Doha, Qatar. e-mail: (smalkuwari@hbku.edu.qa).}, and Ahmed~Farouk\thanks{A.~Farouk is with  the Qatar Center for Quantum Computing, College of Science and Engineering, Hamad Bin Khalifa University, Doha, Qatar and the Department of Computer Science, Faculty of Computers and Artificial Intelligence, Hurghada University, Hurghada, Egypt. e-mail:(ahmed.farouk@sci.svu.edu.eg).}}%

\maketitle

\begin{abstract}
A brain-computer interface (BCI) system enables direct communication between the brain and external devices, offering significant potential for assistive technologies and advanced human-computer interaction. Despite progress, BCI systems face persistent challenges, including signal variability, classification inefficiency, and difficulty adapting to individual users in real time. In this study, we propose a novel hybrid quantum learning model, termed QSVM-QNN, which integrates a Quantum Support Vector Machine (QSVM) with a Quantum Neural Network (QNN), to improve classification accuracy and robustness in EEG-based BCI tasks. Unlike existing models, QSVM-QNN combines the decision boundary capabilities of QSVM with the expressive learning power of QNN, leading to superior generalization performance. The proposed model is evaluated on two benchmark EEG datasets, achieving high accuracies of 0.990 and 0.950, outperforming both classical and standalone quantum models. To demonstrate real-world viability, we further validated the robustness of QNN, QSVM, and QSVM-QNN against six realistic quantum noise models, including bit flip and phase damping. These experiments reveal that QSVM-QNN maintains stable performance under noisy conditions, establishing its applicability for deployment in practical, noisy quantum environments. Beyond BCI, the proposed hybrid quantum architecture is generalizable to other biomedical and time-series classification tasks, offering a scalable and noise-resilient solution for next-generation neurotechnological systems.
\end{abstract}

\begin{IEEEImpStatement}
Integrating quantum computing and machine learning techniques into brain-computer interface (BCI) systems represents a transformative step towards overcoming existing challenges in signal processing and classification accuracy. By harnessing the capabilities of quantum parallelism and entanglement, these innovative approaches have the potential to significantly enhance the efficiency and accuracy of BCI systems, ultimately enabling more seamless interaction between individuals and external devices. The promising results obtained from applying the proposed quantum algorithm, which integrates the quantum neural network (QNN) and the quantum support vector machine (QSVM), to EEG-based BCI datasets demonstrate their superiority over classical methods in terms of accuracy and robustness. 
\end{IEEEImpStatement}

\begin{IEEEkeywords}
Brain-Computer Interface (BCI), ElectroEncephaloGram (EEG), Quantum Computing (QC), Quantum Machine Learning (QML), Quantum Neural Network (QNN), Quantum Support Vector Machine (QSVM)
\end{IEEEkeywords}

%
\IEEEpeerreviewmaketitle

\section{Introduction}\label{Sec1}

A new frontier in human-machine interaction is opened up by brain-computer interface (BCI) systems, which establish a direct connection between the brain and external devices. Electroencephalography (EEG) has emerged as a prominent non-invasive method to obtain brain signals and is at the core of BCI systems \cite{nicolas2012non}. These systems interpret and analyze brain activity and translate it into valuable commands using sophisticated signal processing techniques and decoding algorithms \cite{Maiseli2023}. The role of the user interface is essential to allow interaction and provide customized sensory interfaces, auditory input, and visual displays. Various types of BCI systems exist, such as 
P300-based BCIs, SSVEP BCIs, invasive BCIs, and hybrid BCIs \cite{millan2010combining}. These systems have different applications in areas such as healthcare, neurorehabilitation, neuroprosthetics, assistive technology, gaming, communication devices, and cognitive enhancement \cite{lebedev2006brain}. 

BCI systems face numerous obstacles, necessitating the incorporation of machine learning (ML) methods to improve their performance and dependability. One crucial challenge is the poor signal-to-noise ratio (SNR) in EEG and other neural recording technologies, which makes it difficult to correctly extract meaningful brain activity patterns \cite{Wolpaw2002}. Furthermore, inter- and intra-subject variability in brain signals requires adaptive learning models that generalize between users and conditions \cite{Schalk2004}. Real-time decoding is additionally challenged by the non-stationary nature of brain signals, which alter over time due to fatigue, attention shifts, or environmental influences \cite{Lotte2007}. Furthermore, the high dimensionality of brain data requires feature extraction and dimensionality reduction techniques to increase computational efficiency and classification accuracy \cite{Muller2008}. These issues have resulted in the widespread use of ML techniques such as deep learning (DL) and reinforcement learning (RL). These techniques can improve signal processing and feature extraction and enable more resilient and adaptive BCI systems \cite{Craik2019}.

ML techniques have enabled BCI devices to overcome difficulties by dramatically improving accuracy and efficiency. Traditional classifiers have been frequently used for EEG-based BCI applications due to their simplicity and effectiveness in handling small datasets \cite{lotte2018review}. However, these models struggle with non-linear and multidimensional brain signals. More advanced techniques have shown better performance in automatically extracting significant features from raw EEG signals and capturing temporal dependencies \cite{Schirrmeister2017}. Long short-term memory (LSTM) networks improve real-time decoding by solving the vanishing gradient problem in sequential neural data \cite{Roy2019}. Furthermore, transfer learning approaches have been used to reduce inter-subject variability, allowing models trained on a single group of users to generalize to other individuals \cite{Jayaram2016}. RL has also been investigated for adaptive BCI systems, which allow real-time tweaking of control signals based on user feedback \cite{Perdikis2018}. Incorporation of deep learning (DL) and hybrid ML models continues to improve BCI performance, making it more robust and reliable for real-world applications like neuroprosthetics and assistive technology. Recent studies have demonstrated the effectiveness of deep learning approaches, particularly CNN-based models, for EEG signal classification. For example, the work in \cite{bspc2022_multiscale} proposed a multiscale CNN architecture to enhance spatial spectral feature extraction, achieving high accuracy in motor imagery tasks. Another study \cite{bspc2022_emotion} employed a deep fusion of CNN and GRU for cross-subject emotion recognition, addressing generalization issues in EEG-based systems. Furthermore, in \cite{eaai2022_attention}, a hybrid attention-based CNN-LSTM model was introduced, showing significant improvements in interpretability and classification performance on EEG emotion datasets.

Despite the immense advances achieved by ML in BCI systems, various major obstacles remain, necessitating the investigation of quantum machine learning (QML) as a feasible solution. One key challenge is the large dimensionality and complexity of brain signal data, which makes feature selection and classification computationally intensive for typical machine learning methods \cite{Shenoy2021}.  Furthermore, ML-based BCI models have inadequate subject generalization due to inter-individual heterogeneity in brain activity, resulting in uneven performance between users \cite{Abdulkader2015}. Furthermore, traditional DL algorithms require massive labeled datasets, making them data-hungry and computationally costly during training and adaptation. 
These limitations have driven the potential of QML to process high-dimensional neural data more efficiently than classical methods, leading to increased robustness and reduced computational complexity in BCI applications \cite{Schuld2022}. By utilizing QML, BCI systems could improve their feature extraction, classification, and real-time signal processing performance, making them more adaptive and dependable for neurotechnology applications.

\subsection{Motivation}
\begin{figure*}
    \centering
    \includegraphics[width=\linewidth]{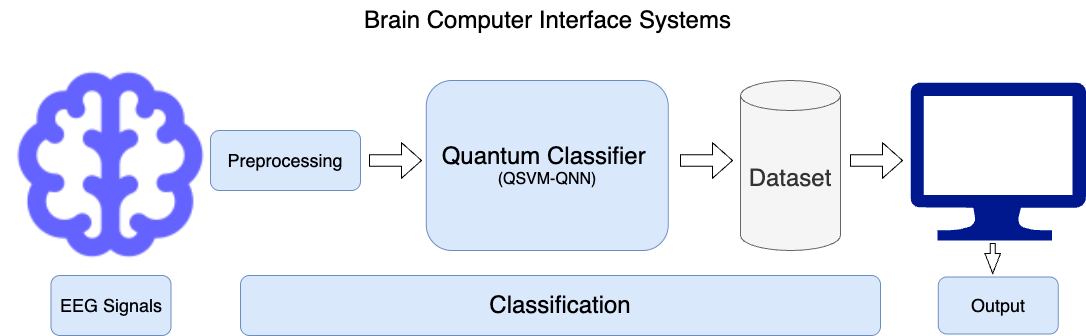}
    \caption{Integration of QML with BCI Systems.}
    \label{fig:schematic}
\end{figure*}

QML has emerged as a feasible approach to solving the problems of classical ML in BCI systems. Several QML algorithms, including Quantum Support Vector Machines (QSVMs) and Quantum Neural Networks (QNNs), have shown potential to overcome the problem of categorization and feature extraction \cite{biamonte2017quantum}. QSVMs use quantum kernel approaches to efficiently process high-dimensional data, whereas QNNs use quantum parallelism to improve pattern recognition and decision-making processes \cite{Schuld2019}. Despite these advances, current QML models frequently lack adaptability and robustness when dealing with non-stationary and noisy inputs, resulting in inferior performance in real-world applications \cite{Lamata2020}. To address this research gap, we propose a hybrid quantum-classical model combining QSVMs and QNNs to improve classification accuracy and adaptability in BCI systems. The QSVM component identifies complicated non-linear patterns from EEG data via quantum-enhanced feature mapping, while the QNN module improves decision-making by utilizing entanglement-based learning representations. The QSVM-QNN model was tested on two EEG-based BCI datasets: the EEG Motor Movement (EEGMM) dataset and the EEG dataset, both containing high-dimensional brain signal data. The model achieved 0.990 accuracy on the EEGMM dataset and 0.950 accuracy on the EEG dataset, outperforming QNN while being comparable to QSVM, demonstrating its potential for robust EEG signal classification in BCI systems. The proposed model was tested against several noise models to ensure its robustness to external disturbances. It showed good resilience to quantum noise, making it viable for real-world neurotechnology applications. However, QSVM-QNN offers a trade-off between accuracy and computational cost, making it suitable for resource-constrained BCI applications. A schematic figure illustrating the implementation of the proposed QSVM-QNN model in BCI systems is shown in Fig. \ref{fig:schematic}.

\subsection{Main Contributions}\label{Sec1.4}
To the best of our knowledge, this is the first work to systematically integrate a QSVM with a QNN in a unified hybrid framework for EEG-based BCI systems. The key technical contributions of this work are summarized as follows:

\begin{itemize}
    \item[1)] We propose a novel hybrid quantum learning framework, termed QSVM-QNN, which leverages the kernel-based discrimination capability of QSVM and the expressive power of QNN for improved EEG signal classification in BCI applications.
    
    \item[2)] We develop and implement an efficient quantum feature encoding strategy adapted to EEG signals. This enables compatibility with both QSVM and QNN circuits and improves feature separability in the quantum latent space.
    
    \item[3)] We conduct a comprehensive evaluation of the proposed QSVM-QNN model against existing classical (e.g., SVM, ANN) and quantum (QNN, QSVM) baselines using two benchmark EEG-based BCI datasets, demonstrating consistent improvement in accuracy and F1 score under noise-free conditions.
    
    \item[4)] We evaluate the robustness and noise resilience of the QSVM-QNN framework under six realistic quantum noise models (bit flip, phase flip, bit-phase flip, amplitude damping, phase damping, and depolarization), highlighting its practicality for deployment in near-term noisy quantum devices.
    
\end{itemize}

\subsection{Organization}
The rest of this paper is organized as follows: Section \ref{Sec2} highlights the fundamentals of quantum computing (QC), QNN, and QSVM. Section \ref{Sec3} describes the overall problem framework and the methodology in detail. Section \ref{Sec4} discusses the experimental results and noisy analysis. Finally, Section \ref{Sec5} concludes the paper.

\section{Background}\label{Sec2}

\subsection{Fundamentals of Quantum Computing}
QC utilizes the principles of quantum mechanics to perform computation. A quantum state \(|\psi\rangle\) can be defined in a complex vector space known as a Hilbert space \(\mathcal{H}\). Quantum bits, or qubits, are the building blocks of quantum information. The ability of qubits to exist in superpositions of states \cite{schrodinger1935} is a key principle \cite{bohr1913}.
It can exist in a superposition of states $\ket{0}$ and $\ket{1}$ simultaneously be represented as:
\begin{eqnarray}
    |\psi\rangle = \alpha|0\rangle + \beta|1\rangle,
\end{eqnarray}
where \(\alpha\) and \(\beta\) are probability amplitudes satisfying \(\left|\alpha\right|^2 + \left|\beta\right|^2 = 1\). For $n$ qubits, a quantum state can represent $2^n$ classical states, providing inherent parallelism. Entanglement, famously described in the EPR paradox paper \cite{einstein1935}, is a fundamental quantum phenomenon in which qubits exhibit strong correlations in ways that classical systems cannot emulate. An entangled state can be represented as:
\begin{eqnarray}
    |\psi\rangle = \frac{1}{\sqrt{2}}(|00\rangle + |11\rangle).
\end{eqnarray}
Quantum gates \cite{deutsch1985}, analogous to classical logic gates, manipulate qubit states. The transformation of a quantum state through gates is governed by unitary operators. The action of a Hadamard gate, for instance, can be expressed as:
\begin{eqnarray}
    H|0\rangle = \frac{1}{\sqrt{2}}(|0\rangle + |1\rangle).    
\end{eqnarray}
Quantum measurement projects a qubit's state onto the classical basis. The measurement outcome is probabilistic. For instance, measuring in the computational basis yields:
\begin{eqnarray}
    |0\rangle \text{ or } |1\rangle    
\end{eqnarray}
with respective probabilities \(\left|\alpha\right|^2\) and \(\left|\beta\right|^2\). Pioneering work \cite{shor1994} and \cite{grover1996} introduced quantum algorithms that exploit quantum phenomena for specific computational advantages.

\subsection{Fundamentals of Quantum Neural Network (QNN)}

QNNs combine principles from quantum computing and classical artificial neural networks to model complex relationships in data using quantum-mechanical properties such as superposition and entanglement \cite{schuld2018circuit}. In QNNs, information is encoded using qubits, which can exist in a superposition of states, enabling parallel processing of multiple computational paths.
Each quantum neuron receives classical input features \(x_1, x_2, \ldots, x_n\), which are first encoded into quantum states. The processing within a quantum neuron can be abstractly represented as:
\begin{eqnarray}
    |x_1, x_2, \ldots, x_n\rangle \rightarrow |y\rangle,
\end{eqnarray}
where \(|x_1, x_2, \ldots, x_n\rangle\) denotes the input quantum state and \(|y\rangle\) is the quantum output state.
To introduce non-linearity, QNNs may employ quantum versions of activation functions. For instance, the sigmoid function is defined as:
\begin{eqnarray}
    \text{sigmoid}(x) = \frac{1}{1 + e^{-x}},
\end{eqnarray}
where \(x\) is the weighted input to the quantum neuron. In a quantum context, such functions are simulated through parameterized quantum circuits that modulate gate operations accordingly.
A QNN typically consists of multiple layers. The output of one layer forms the input for the next. This transformation can be generalized as:
\begin{eqnarray}
    |x^{(l)}_1, x^{(l)}_2, \ldots, x^{(l)}_n\rangle \rightarrow |y^{(l)}_1, y^{(l)}_2, \ldots, y^{(l)}_m\rangle,
\end{eqnarray}
where \(x^{(l)}_i\) represents the \(i^\text{th}\) input qubit to layer \(l\), and \(y^{(l)}_j\) is the \(j^\text{th}\) output qubit from the same layer. Variational quantum circuits (VQCs) are typically used to implement transformations between these layers, and their parameters are optimized via classical gradient-based methods.

\subsection{Fundamentals of Quantum Support Vector Machine (QSVM)}

QSVMs extend classical SVMs by utilizing quantum mechanics to perform classification tasks in high-dimensional quantum feature spaces \cite{schuld2018quantum}. In QSVM, classical data points \(x_i \in \mathbb{R}^d\) are encoded into quantum states via a quantum feature map \(\phi\), yielding:
\begin{eqnarray}
    |x_i\rangle \rightarrow |\phi(x_i)\rangle,
\end{eqnarray}
where \(|\phi(x_i)\rangle\) is the quantum state representing the feature-embedded input.
The key advantage of QSVM lies in its ability to evaluate quantum kernels, defined as:
\begin{eqnarray}
    K(x_i, x_j) = \langle\phi(x_i)|\phi(x_j)\rangle,
\end{eqnarray}
which represents the inner product between two quantum states in the quantum feature space. This operation enables the computation of similarities between input samples more efficiently than classical kernels in certain regimes.
The QSVM aims to find a decision boundary expressed as:
\begin{eqnarray}
    f(x) = \sum_{i} \alpha_i K(x_i, x),
\end{eqnarray}
where \(x\) is the test sample, \(x_i\) are the training samples (support vectors), \(\alpha_i\) are learned coefficients representing the contribution of each support vector, and \(K(x_i, x)\) is the quantum kernel.
To determine the optimal parameters \(\alpha_i\), QSVM minimizes the quantum version of the hinge loss function:
\begin{eqnarray}
    L(y, f(x)) = \max(0, 1 - y \cdot f(x)),
\end{eqnarray}
where \(y \in \{-1, +1\}\) is the true class label for sample \(x\), and \(f(x)\) is the predicted score. The loss penalizes incorrect classifications and samples that fall within the decision margin.
The QSVM framework uses quantum circuits to compute the kernel and apply entangling gates to capture complex correlations between input features \cite{lloyd2013quantum, rebentrost2014quantum}, offering a scalable alternative to classical SVMs for high-dimensional tasks.

\section{Methodology}\label{Sec3}
\subsection{Problem Formulation and Overall Framework}
Consider a set of data points represented by a collection of feature vectors $X = \{\vec{x}_1, \vec{x}_2, \ldots, \vec{x}_n\}$, where each feature vector $\vec{x}_i \in \mathbb{R}^d$ represents a data point with d features. The goal is to classify the dataset into two classes, making it a binary classification problem. Here, three classification models are used. The first is QNN, which is trained using a training dataset consisting of feature vectors and their corresponding labels. The model learns a mapping function $f: \mathbb{R}^d \rightarrow \{0,1\}$ that maps a feature vector $x_i$ to its corresponding label $y_i$. The model's performance is evaluated on a separate test dataset, and its accuracy is measured as the proportion of correctly classified samples. Secondly, for QSVM, a k-means clustering method needs to be implemented to find two centroids. Then, the classes of the data points are determined by comparing the inner products between the centroid and the test data points. Finally, the proposed QSVM-QNN comprises QSVM and QNN together in one quantum circuit, which starts with the QSVM part and then adds the QNN layer to learn the optimal parameters after minimizing the cost function. The mathematical representation of the three models is as follows:

\begin{eqnarray}
\hat y &=& F ( \hat M [\text{QNN}\{\ket{\vec{x}}  \}     ] ),\nonumber\\
\hat y &=& F ( \hat M [\text{QSVM}\{\ket{\vec{x}}  \}     ] ),\nonumber\\
\hat y &=& F ( \hat M [\text{QSVM-QNN}\{\ket{\vec{x}}  \}     ] )
\end{eqnarray}

where $\ket{\vec{x}} = Encoding(X_{PCA})$, $\hat M$ is the measurement in the quantum circuit and $F$ represents the function output.

\subsection{Quantum Neural Network (QNN)}
The essential component of a QNN is a quantum neuron, which processes classical inputs and generates a quantum state as an output. The structure of a quantum neuron consists of three primary layers: encoding, quantum processing, and measurement. The encoding layer maps classical input features $\vec{x}$ to a quantum state $|\psi\rangle$ using quantum feature maps. One common approach for encoding classical data into quantum states is amplitude encoding, where the classical input $\vec{x}$ is mapped into a normalized quantum state as follows:
\begin{equation}
    |\psi\rangle = \sum_{i} \sqrt{p_i} |i\rangle
\end{equation}
The quantum processing layer applies a parameterized unitary operation $V(\mathbf{\theta})$ to the quantum state:
\begin{equation}
    V(\mathbf{\theta})|\psi\rangle = \sum_i \sqrt{p_i} V(\mathbf{\theta}) |i\rangle
\end{equation}
The measurement layer measures on a computational basis, mapping the quantum state to classical probabilities:
\begin{equation}
    \text{Pr}(c_i) = \left|\langle c_i | V(\mathbf{\theta}) |\psi\rangle\right|^2
\end{equation}
The quantum neuron function takes classical inputs $\vec{x}$, encodes parameters $\mathbf{\theta}$, and returns the output probabilities $\text{Pr}(c_i)$ after the measurement:
\begin{equation}
    \text{Pr}(c_i) = \left|\langle c_i | V(\mathbf{\theta}) \sum_i \sqrt{p_i} |i\rangle\right|^2
\end{equation}
In a quantum layer with $N$ neurons, the combined quantum state is given by:
\begin{equation}
    |\psi_{\text{layer}}\rangle = \bigotimes_{i=1}^{N} V_i(\mathbf{\theta}_i)|\psi_i\rangle
\end{equation}
\begin{figure*}[]
    \centering
    \includegraphics[width=\linewidth]{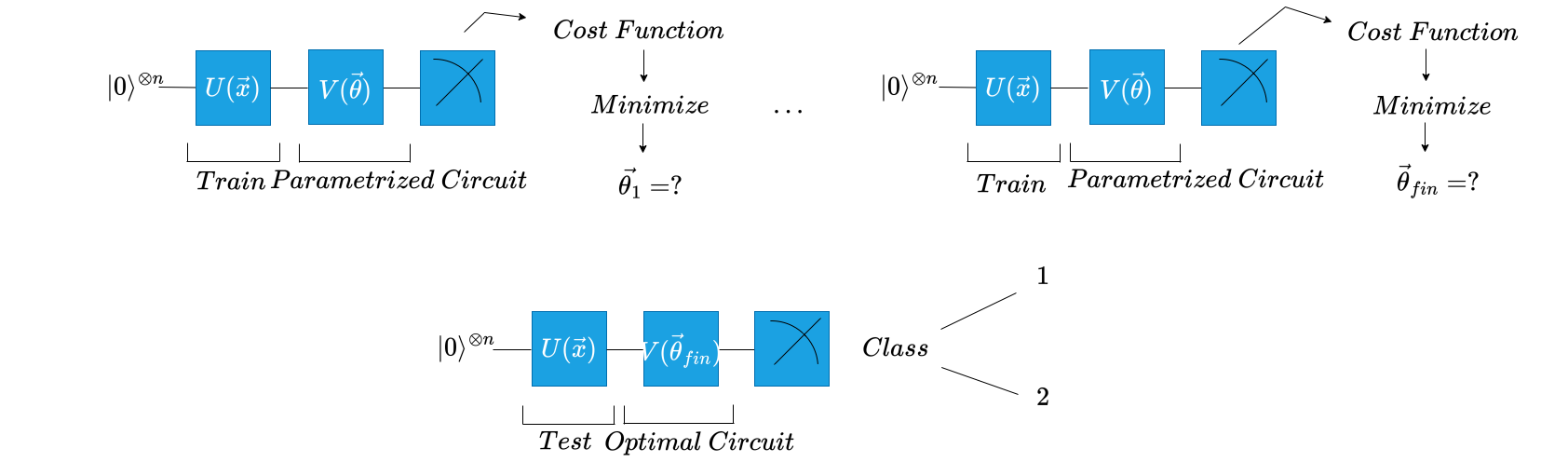}
    \caption{Quantum Circuit of QNNs Process.}
    \label{fig:enter-label-2}
\end{figure*}
\begin{algorithm}
\caption{Quantum Neural Network (QNN)}
\label{Algo-1}
\textbf{Input:} Dataset \\
\textbf{Output:} Evaluation Metrics
\begin{algorithmic}[1]
\State Import necessary libraries and modules
\State Load and preprocess the dataset

\Function{QuantumCircuit}{$\text{parameters}, {x}$}
    \State Create a quantum circuit
    \State Apply quantum gates based on parameters and features
    \State Measure the output qubit 
    \State \Return \texttt{qc}
\EndFunction

\Function{CostFunction}{$\text{parameters}, X, y$}
    \State Initialize \texttt{total\_cost} to $0$
    \For{$i$ in range(\texttt{len(X)})}
        \State $x \gets X.iloc[i].\text{values}$
        \State \texttt{qc = QuantumCircuit(parameters, x)}
        \State Evaluate \texttt{qc} and update \texttt{total\_cost}
    \EndFor
    \State \Return \texttt{total\_cost}
\EndFunction

   \State Choose $epochs$ value
   \State $initial\_parameters \gets$ Randomly initialized parameters
   \For{$k = 0$ to $epochs-1$}
       \State $optimize \gets$ Minimize cost function using COBYLA method
       \State $initial\_parameters \gets optimal\_parameters$
   \EndFor

\State Evaluate the trained model on the test set
\State {Calculate evaluation metrics}
\State {Display the results}

\end{algorithmic}
\end{algorithm}

A QNN consists of multiple quantum layers, each followed by classical processing layers. The final output is obtained by applying a classical readout layer to the measured quantum states:
\begin{equation}
    \text{Output} = \mathbf{W} \cdot \left[ \text{Pr}(c_{i,1}), \text{Pr}(c_{i,2}), \ldots, \text{Pr}(c_{i,N}) \right]^T
\end{equation}
Here $\mathbf{W}$ is the weight matrix in the classical readout layer. The training of a QNN requires optimizing the encoding parameters $\vec{\theta}$ and the classical readout weights $\mathbf{W}$ to minimize a chosen cost function:
\begin{equation}
    \text{Cost}(\mathbf{\theta}, \mathbf{W}) = \sum_{\text{training samples}} \left( \text{Predicted} - \text{Target} \right)^2
\end{equation}

This involves a gradient-free optimization technique using the COBYLA optimizer. Fig. \ref{fig:enter-label-2} and the Algorithm \ref{Algo-1} give the quantum circuit and the step-by-step process.

\subsection{Quantum Support Vector Machine (QSVM)}

\begin{figure}
    \centering
    \includegraphics[width=\linewidth]{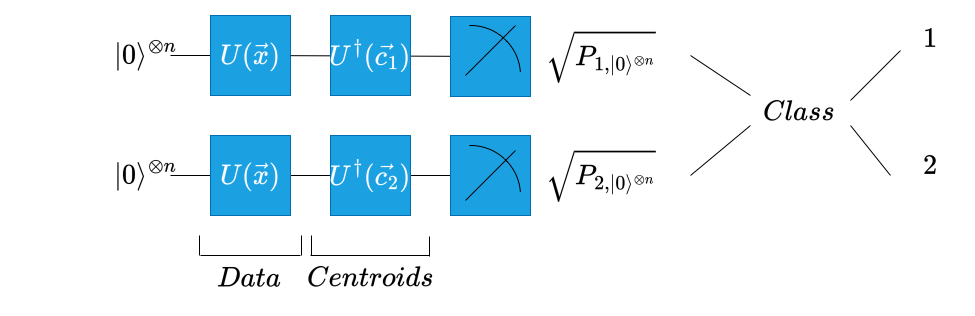}
    \caption{QSVM Quantum Circuit for Classification.}
    \label{fig:enter-label-3}
\end{figure}

Here, we explain the QSVM algorithm, particularly following the $UU^\dagger$ method as described in \cite{bib_Kariya}. The primary objective of this algorithm is to compute the inner product between the centroid of a dataset cluster and a given test data point in the quantum feature space. This computation uses a quantum circuit, as illustrated in Fig. \ref{fig:enter-label-3} and given by Algorithm 2. The mathematical formulation of the QSVM is detailed as follows: the encoding of data can be written as:
\begin{eqnarray}
\ket{\psi} = U(\vec{x})\ket{0}^{\otimes n}    
\end{eqnarray}

The encoding of two centroid data points into quantum states is made as follows:

\begin{eqnarray}
\ket{\psi_1} &=& U(\Vec{c_1})\ket{0}^{\otimes n},\ \ket{\psi_2} = U(\Vec{c_2})\ket{0}^{\otimes n}
\label{Eq1}
\end{eqnarray}

The two inner products between the data and the two centroids are expressed by:
\begin{eqnarray}
\braket{\psi_1|\psi} &=& ^{n \otimes}\bra{0}U(\Vec{c_1})^{\dagger}U(\vec{x})\ket{0}^{\otimes n}= ^{n \otimes}\bra{0}A_1\ket{0}^{\otimes n},\label{eq_3}\nonumber\\ 
\braket{\psi_2|\psi} &=& ^{n \otimes}\bra{0}U(\Vec{c_2})^{\dagger}U(\vec{x})\ket{0}^{\otimes n}= ^{n \otimes}\bra{0}A_2\ket{0}^{\otimes n} \nonumber\\
\label{eq_4}
\end{eqnarray}
\begin{algorithm}
\caption{Quantum Support Vector Machine (QSVM)}
\label{algo-qsvm}
\textbf{Input:} Dataset \\
\textbf{Output:} Evaluation Metrics
\begin{algorithmic}[1]

\State {Import} necessary libraries and modules
\State Load and preprocess the dataset

\State{Apply KMeans clustering on the dataset features}
\State{Obtain cluster labels and centroids}

\For{$i \gets 0$ to $len(X\_test)-1$}
    \State{Encode the test data in $U$ operation}
    \State{Encode the centroid data in $U^{\dagger}$ operation}
    \State{Run quantum circuits and obtain probabilities for both centroids}
    \State{Assign predicted label based on probabilities}
\EndFor

\State{Calculate evaluation metrics}

\State{Display results}
\end{algorithmic}
\end{algorithm}

This signifies the operations \(A_1 = U(\Vec{c_1})^{\dagger}U(\vec{x}),A_2 = U(\Vec{c_2})^{\dagger}U(\vec{x})\), where \(A_1, A_2\) are arbitrary \(n\)-qubit operators that produce arbitrary states in the \(n\)-qubit system when applied to the state \(\ket{0}^{\otimes n}\). The states \(A_1\ket{0}^{\otimes n},A_2\ket{0}^{\otimes n}\) have the following general form:
\begin{eqnarray}
A_1\ket{0}^{\otimes n} &=& \alpha_{1,0}\ket{000...000} + \alpha_{1,1}\ket{000...001} \nonumber\\
&+& \ldots + \alpha_{1,2^m-1}\ket{111...111},\\
A_2\ket{0}^{\otimes n} &=& \alpha_{2,0}\ket{000...000} + \alpha_{2,1}\ket{000...001} \nonumber\\
&+& \ldots + \alpha_{2,2^m-1}\ket{111...111}
\label{4}
\end{eqnarray}

Eq. \eqref{eq_4} can be simplified to:
\begin{eqnarray}
^{n \otimes}\bra{0}A_1\ket{0}^{\otimes n} &=& \alpha_{1,0}, \ ^{n \otimes}\bra{0}A_2\ket{0}^{\otimes n} = \alpha_{2,0}
\label{5}
\end{eqnarray}

Here \( \alpha_{1,0}, \alpha_{2,0}\) are the coefficients of \(\ket{000...000}\), real numbers \cite{Shradha2023} representing the square root of the probability \(P_{000...000}\) of measuring \(\ket{000...000}\). This is expressed as:
\begin{eqnarray}
\alpha_{1,0} &=& \sqrt{P_{1,000...000}}= \sqrt{P_{1,\ket{0}^{\otimes n}}}\label{28}\\
\alpha_{2,0} &=& \sqrt{P_{2,000...000}}=\sqrt{P_{2,\ket{0}^{\otimes n}}}
\label{6}
\end{eqnarray}

Combining Eqs. \eqref{eq_4}, \eqref{28}, and \eqref{6}, we have:
\begin{eqnarray}
\braket{\psi_1|\psi} &=& \sqrt{P_{1,\ket{0}^{\otimes n}}},\ \braket{\psi_2|\psi} = \sqrt{P_{2,\ket{0}^{\otimes n}}}
\label{7}
\end{eqnarray}

\begin{figure*}
    \centering
    \includegraphics[width=\linewidth]{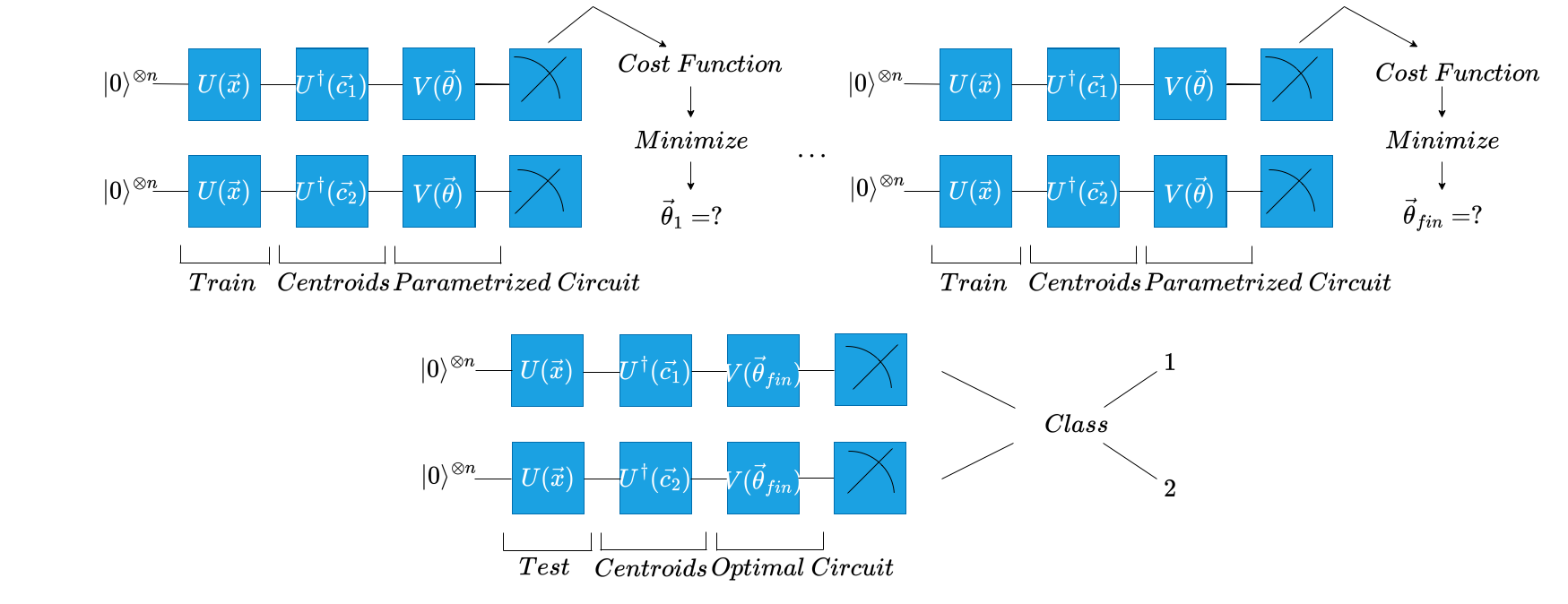}
    \caption{Quantum Circuit of QSVM-QNN Process.}
    \label{fig:enter-label-4}
\end{figure*}
\subsection{The Proposed QSVM-QNN}
Here, we propose a novel QNN learning algorithm based on the QSVM method.  The proposed QSVM-QNN integrates a modified QNN with a Variational Quantum Circuit (VQC), incorporating tunable parameters ($\vec{\theta}$) for enhanced learning capabilities. The proposed QSVM-QNN algorithm consists of two key components, the $UU^{\dagger}$ method and VQC, as shown in Fig. \ref{fig:enter-label-4}. The $UU^{\dagger}$ method follows the same encoding process here, where data points and centroids are mapped to quantum states using $U$ and $U^{\dagger}$ operations. After the $U^{\dagger}$ operation, the VQC is applied to the circuit, consisting of a sequence of single-qubit rotation operators such as $R_x(\theta)$, $R_y(\theta)$ and $R_z(\theta)$, and two-qubit controlled-Not ($CX$) gates. These rotation operators have angle parameters, which are tuned during optimization to find the optimal parameters in the quantum circuit. The step-by-step process of QSVM-QNN is given in Algorithm \ref{algo-qsvm-qnn-learning}. The rotation operators and $CX$ are defined as follows:

\begin{algorithm}
\caption{QSVM based QNN (QSVM-QNN)}
\label{algo-qsvm-qnn-learning}
\textbf{Input:} Dataset \\
\textbf{Output:} Evaluation Metrics
\begin{algorithmic}[1]

\State {Import} necessary libraries and modules
\State Load and preprocess the dataset

\State{Apply KMeans clustering on the Dataset features}
\State{Obtain cluster labels and centroids}

\State{Define a parametrized quantum circuit}

\State{Define a cost function for optimization}

\State{Initialize parameters randomly}

\For{$iteration \gets 1$ to $num\_iterations$}
    \State{Optimize the cost function parameters using COBYLA method}
    \State{parameters $\gets$ optimal\_parameters}
\EndFor

\For{$i \gets 0$ to $len(X\_test)-1$}
    \State{Create quantum circuits for $U$ and $U^\dagger$ operations using optimal\_parameters}
    \State{Run quantum circuits and obtain probabilities}
    \State{Assign predicted label based on probabilities}
\EndFor

\State{Calculate evaluation metrics}

\State{Display results}
\end{algorithmic}
\end{algorithm}

\begin{eqnarray}
R_{x}(\theta) &=&\begin{bmatrix}
    \cos{\frac{\theta}{2}}& -i\sin{\frac{\theta}{2}}\\
    -i \sin{\frac{\theta}{2}}& \cos{\frac{\theta}{2}}
\end{bmatrix},
R_{y}(\theta) =\begin{bmatrix}
    \cos{\frac{\theta}{2}}& -\sin{\frac{\theta}{2}}\\
    \sin{\frac{\theta}{2}}& \cos{\frac{\theta}{2}}
\end{bmatrix},\nonumber\\
R_{z}(\theta) &=&\begin{bmatrix}1& 0\\ 0 & e^{i\theta}\end{bmatrix},
CX =\begin{bmatrix}I & O\\ O & X\end{bmatrix}
\label{8}
\end{eqnarray}
where $I, O$ and $X$ are the identity, zero, and Pauli X matrices, respectively. For the experiment, the VQC circuit for one-layer is defined as,

\begin{eqnarray}
\text{VQC}&\equiv& R_x({\theta}_1)_1R_x({\theta}_2)_2...R_x({\theta}_n)_nCX_{1,2}...CX_{n-1,n}\nonumber\\
&&R_y({\theta}_1)_1R_y({\theta}_2)_2...R_y({\theta}_n)_nCX_{1,2}...CX_{n-1,n}\nonumber\\
&&R_z({\theta}_1)_1R_z({\theta}_2)_2...R_z({\theta}_n)_nCX_{1,2}...CX_{n-1,n}\nonumber\\
\end{eqnarray}

Similarly to the QNN process, the quantum circuit is run, and the optimized parameters are found, which are then used as the initial parameters of the next QNN layer. After finding the final optimal parameters, they are used in the test dataset circuit to classify them into two classes.

To summarize, the proposed architecture comprises two key components:
\begin{itemize}
        \item {QSVM Module:} Implements the $UU^{\dagger}$ method to encode the data and compute the inner products between the quantum states corresponding to data points and centroids. This is achieved by sequentially applying the $U$ and $U^{\dagger}$ unitary operators.
        \item {QNN Module:} A VQC is employed and composed of:
        \begin{itemize}
            \item Parameterized single-qubit gates: $R_x(\theta)$, $R_y(\theta)$, and $R_z(\theta)$,
            \item Two-qubit entangling gates: Controlled-NOT (CX) gates in a layered entanglement structure.
        \end{itemize}
\end{itemize}
The total number of trainable parameters in the VQC is $\mathcal{O}(n \cdot d)$, where $n$ is the number of qubits and $d$ is the number of variational layers in the circuit.
The computational complexity of the model is as follows: \begin{itemize}
        \item The QSVM module requires a constant-depth quantum circuit to perform the overlap (inner product) operation between the quantum states.
        \item The QNN module contributes polynomial complexity, i.e. $\mathcal{O}(n \cdot d)$ in terms of gate depth and parameter count.
    \end{itemize}

\section{Experimental Results}\label{Sec4}

\subsection{Hyperparameters and Metrics}\label{Sec4.1}
All the quantum circuits used in the experiment are run using the Aer backend simulator, named `automatic'. The noisy results are collected using Aer's noisy simulator. All circuits are executed using a total of 1024 shots. For QNN and QSVM-QNN, one layer of parametrized circuit is used, while 5 epochs of QNN are used to collect the optimal parameters to classify the test data points. For all the classical and quantum algorithms, the test size is taken to be 0.2 for both datasets. For quantum algorithms, 3 qubits are used. In the noisy simulation, all the noise parameters are chosen to range from 0 to 1 to estimate the accuracy. A maximum number of data points is taken for the algorithms to run, while for noisy simulation, 100 data points are chosen to collect the results faster. Four metrics, accuracy, precision, recall, and F1 measure, are used to evaluate classical and quantum algorithms.

\subsection{Datasets}
The EEG Motor Movement (EEGMM) dataset, available on Kaggle \cite{eeg_motor_movement}, is a collection of EEG recordings during motor movement and imagery tasks. The dataset captures brain activity through EEG signals as subjects perform various motor-related activities. It consists of 9760 data points and 65 columns. It has features such as time, Fc5, Fc3, Fc1,..., O2, and Iz. The EEG dataset, accessible on Kaggle \cite{eeg_dataset}, comprises recordings of EEG signals obtained from human subjects. This dataset includes information on various cognitive states, sensory stimuli, or tasks performed by subjects during the recording sessions. It consists of 8064 data points and 32 columns. It has features such as Fp1, AF3, F3,..., PO4, and O2.

\subsection{Preprocessing}
The methodology starts by systematically preprocessing, such as understanding and evaluating the dataset features, including identifying the relevant feature columns and removing irrelevant ones. This is achieved using the Principal Component Analysis (PCA) method. The number of features is chosen to encode the data points properly into the amplitude encoding circuit. In this case, a binary classification approach is performed, classifying the dataset into two classes. In the case of QSVM and QSVM-QNN, k-means clustering is performed to divide the dataset into two clusters and find two centroids, which are then encoded into the quantum circuits for further processing.

\subsection{Comparative Models}
Both classical and quantum algorithms are used as baseline models to evaluate the proposed QSVM-QNN algorithm. The KNN classifier is employed to augment instances by considering the majority class of their KNNs, providing a non-parametric approach. The RF classifier evaluates overall predictive performance and mitigates overfitting risks. A SVM classifier with a linear kernel effectively partitions high-dimensional spaces. With an MLP classifier and a hidden layer, CNN offers a sophisticated nonlinear classification approach. An LR classifier with a multinomial solver provides a linear classification approach. The DT classifier partitions the feature space into regions to establish interpretable decision boundaries. An NB classifier with a multinomial distribution provides a probabilistic viewpoint for simpler feature spaces. 

\subsection{Noise Models}
The quantum algorithms are tested for robustness against six different noise models, including bit flip (BF), phase flip (FP), bit phase flip (BPF), amplitude damping (AD), phase damping (PD), and depolarization (DP), as described in \cite{bib_Kumar}. In quantum systems, these noise models are often encountered. They play a crucial role in the development of error-correction strategies for reliable quantum computation, in understanding and mitigating noise effects in quantum circuits and algorithms, and in offering insights into the behavior of quantum devices.

\subsection{Results and Analysis}\label{Sec4.3}
The two datasets, EEGMM and EEG, are used to test the accuracy of the proposed QSVM-QNN, quantum techniques such as QNN and QSVM, and classical algorithms such as KNN, RF, SVM, CNN, LR, DT, and NB. Four metrics are computed for each of the aforementioned algorithms, as shown in Table \ref{table:performance_metrics}. In the EEG data set, SVM and LR had the highest scores (0.999) compared to other classical methods. For the EEGMM dataset, LR has scored the highest accuracy of 0.997 among other classical algorithms. In general, LR achieves the highest accuracy among other classical algorithms. According to the precision metric, LR achieves the highest score of 0.997, while for recall, RF, SVM, and LR achieve a 1.000 score, and for the F1 measure, SVM achieves the highest score of 0.999. In the case of quantum algorithms, QSVM-QNN has the highest accuracy of 0.997 on the EEGMM dataset, while QSVM has the highest accuracy of 0.998 on the EEG dataset, with QSVM-QNN following second at 0.950. It can be noted that, for 100 data points, the majority of classical algorithms, QSVM and QSVM-QNN, achieve 1.000 accuracy; however, their accuracy varies when the maximum number of data points is taken. To collect the findings in Table \ref{table:performance_metrics}, a maximum number of data points is taken for all algorithms, resulting in differences in accuracy of less than 1.000.
Among all classical and quantum algorithms, QNN has the lowest and highest accuracy in the EEG dataset compared to the EEGMM dataset. Furthermore, QNN has shown precision, recall, and an F1 measure of 0.000, making it the worst model for comparison. QSVM and QSVM-QNN provide similar metrics in the two datasets.

\begin{figure}
    \centering
    \includegraphics[width=\linewidth]{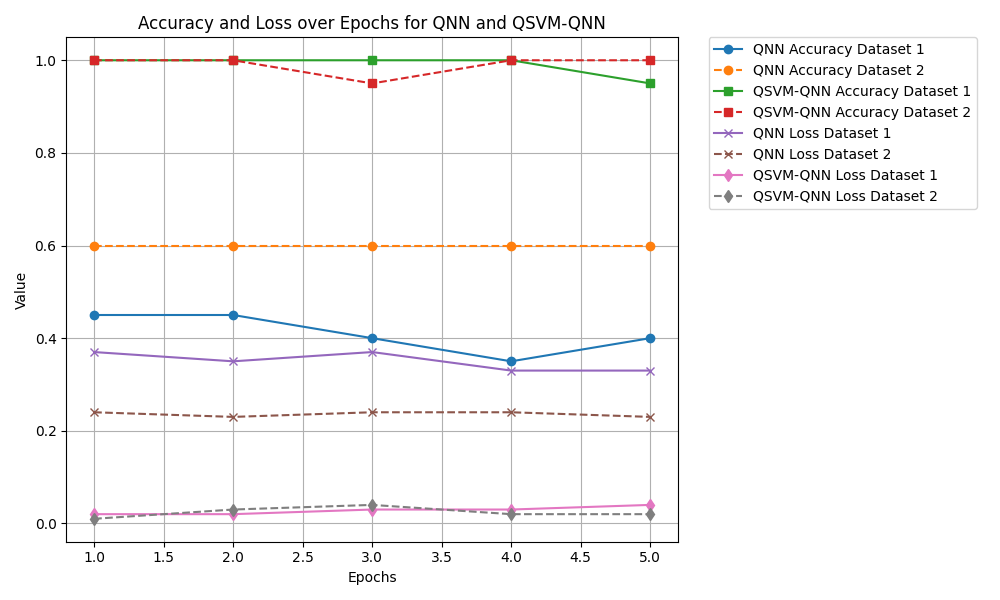}
    \caption{Accuracy and Loss for QNN and QSVM-QNN on EEGMM and EEG Datasets.}
    \label{fig5:}
\end{figure}

According to the precision metric, QSVM has shown a higher score of 0.990 for the EEGMM dataset compared to the QSVM-QNN algorithm, while, in the case of recall, QSVM-QNN has achieved the highest score of 1.000 on the EEG dataset. Following this, QSVM and QSVM-QNN have similar F1 measure values of 0.989 and 0.988 in the EEGMM dataset. Furthermore, Table \ref{table:accuracy_comparison} provides a further analysis of the EEG dataset between our algorithms and the existing work, where a maximum of 0.979 accuracy is achieved using the XGBoost algorithm on the EEGEyeNet dataset. It can be easily realized that the proposed QSVM-QNN algorithm, which shows a maximum accuracy of 0.990 on the EEGMM dataset, is a promising algorithm for implementation in similar datasets. Fig. \ref{fig5:} shows the accuracy and loss versus the number of iterations/epochs of the QNN and QSVM-QNN algorithms on the EEGMM and EEG datasets. For the EEGMM dataset, QNN's accuracy decreases with an increase in epoch and then remains constant over the next epochs. In contrast, QSVM-QNN exhibits steady accuracy across all epochs except for epoch 3, where it degrades slightly. For the EEG dataset, QNN shows constant accuracy with no fluctuation, but QSVM-QNN exhibits consistent accuracy after initially scaling up from a lower value. QSVM-QNN loss is usually close to zero throughout the epochs, whereas the QNN loss ranges from 0.25 to 0.30.

\begin{table*}[ht]
\centering
\caption{Performance Metrics of Classical and Quantum Algorithms for EEGMM and EEG Datasets}
\begin{tabular}{l l c c c c}
\toprule
\textbf{Algorithm} & \textbf{Dataset} & \textbf{Accuracy} & \textbf{Precision} & \textbf{Recall} & \textbf{F1 Measure} \\
\midrule
\multicolumn{6}{l}{\textbf{Classical Classification Algorithms}} \\
\midrule
KNN         & EEGMM & 0.979 & 0.979 & 0.981 & 0.980 \\
KNN         & EEG   & 0.998 & 0.989 & 0.984 & 0.978 \\
RF          & EEGMM & 0.977 & 0.978 & 0.978 & 0.978 \\
RF          & EEG   & 0.999 & 0.989 & 1.000 & 0.995 \\
SVM         & EEGMM & 0.995 & 0.992 & 0.995 & 0.999 \\
SVM         & EEG   & 0.999 & 0.989 & 1.000 & 0.995 \\
CNN         & EEGMM & 0.996 & 0.996 & 0.998 & 0.997 \\
CNN         & EEG   & 0.998 & 0.968 & 0.989 & 0.979 \\
LR          & EEGMM & 0.997 & 0.997 & 0.998 & 0.998 \\
LR          & EEG   & 0.999 & 0.989 & 1.000 & 0.995 \\
DT          & EEGMM & 0.932 & 0.938 & 0.932 & 0.935 \\
DT          & EEG   & 0.997 & 0.968 & 0.978 & 0.973 \\
NB          & EEGMM & 0.902 & 0.914 & 0.897 & 0.906 \\
NB          & EEG   & 0.967 & 0.638 & 0.968 & 0.769 \\
\midrule
\multicolumn{6}{l}{\textbf{Quantum Classification Algorithms}} \\
\midrule
QNN         & EEGMM & 0.450 & 0.667 & 0.429 & 0.522 \\
QNN         & EEG   & 0.600 & 0.000 & 0.000 & 0.000 \\
QSVM        & EEGMM & 0.987 & 0.990 & 0.987 & 0.989 \\
QSVM        & EEG   & 0.998 & 0.989 & 0.967 & 0.989 \\
QSVM-QNN    & EEGMM & 0.990 & 0.988 & 0.988 & 0.988 \\
QSVM-QNN    & EEG   & 0.950 & 0.889 & 1.000 & 0.941 \\
\bottomrule
\end{tabular}
\label{table:performance_metrics}
\end{table*}

\subsection{Statistical Significance Tests}

To assess the statistical validity of performance differences between the evaluated algorithms, we performed one-way ANOVA and independent sample t-tests using the metrics reported in Table~\ref{table:performance_metrics}. Specifically, a one-way ANOVA was performed independently for Accuracy, Precision, Recall, and F1 Score across all models. The results are summarized in Table~\ref{tab:anova_all_metrics}. The ANOVA analysis confirms that the differences in mean values between the algorithms are highly statistically significant for all four metrics (e.g., $p = 1.47 \times 10^{-195}$ for Accuracy). Following ANOVA, we conducted pairwise independent sample t-tests to evaluate whether the proposed hybrid quantum model (QSVM-QNN) statistically outperforms each baseline method. Tables~\ref{tab:ttest_accuracy}, \ref{tab:ttest_precision}, \ref{tab:ttest_recall}, and \ref{tab:ttest_f1} detail the t-statistics and p-values for Accuracy, Precision, Recall, and F1 Score, respectively. The t-test results demonstrate that QSVM-QNN significantly outperforms classical algorithms such as DT, NB, and QNN in all evaluation metrics ($p < 0.05$). Particularly large t-statistics for NB and QNN highlight consistent performance advantages. Conversely, comparisons with QSVM yield non-significant p-values across all metrics, suggesting that QSVM-QNN and QSVM offer statistically similar performance. These findings reinforce the effectiveness of hybrid quantum approaches while also validating the stability and robustness of our proposed QSVM-QNN model.

\begin{table}[ht]
\centering
\caption{ANOVA Test Results for Performance Metrics}
\begin{tabular}{lcc}
\toprule
\textbf{Metric} & \textbf{F-statistic} & \textbf{p-value} \\
\midrule
Accuracy  & 272997.00 & 1.47e-195 \\
Precision & 483358.23 & 7.14e-210 \\
Recall    & 407942.60 & 2.22e-205 \\
F1 Score  & 316065.82 & 4.03e-198 \\
\bottomrule
\end{tabular}
\label{tab:anova_all_metrics}
\end{table}

\begin{table}[ht]
\centering
\caption{T-Test Results: Accuracy (QSVM-QNN vs Others)}
\begin{tabular}{lcc}
\toprule
\textbf{Compared With} & \textbf{t-statistic} & \textbf{p-value} \\
\midrule
KNN & -45.34 & 5.21e-20 \\
RF & -43.26 & 1.20e-19 \\
SVM & -69.96 & 2.21e-23 \\
CNN & -62.18 & 1.83e-22 \\
LR & -70.12 & 2.12e-23 \\
DT & 10.79 & 2.73e-09 \\
NB & 68.98 & 2.85e-23 \\
QNN & 1029.57 & 2.18e-44 \\
QSVM & -56.01 & 1.19e-21 \\
\bottomrule
\end{tabular}
\label{tab:ttest_accuracy}
\end{table}

\begin{table}[ht]
\centering
\caption{T-Test Results: Precision (QSVM-QNN vs Others)}
\begin{tabular}{lcc}
\toprule
\textbf{Compared With} & \textbf{t-statistic} & \textbf{p-value} \\
\midrule
KNN & -107.58 & 9.75e-27 \\
RF & -100.96 & 3.05e-26 \\
SVM & -119.17 & 1.55e-27 \\
CNN & -103.58 & 1.93e-26 \\
LR & -123.36 & 8.32e-28 \\
DT & -25.21 & 1.71e-15 \\
NB & 442.82 & 8.58e-38 \\
QNN & 1477.03 & 3.29e-47 \\
QSVM & -141.09 & 7.44e-29 \\
\bottomrule
\end{tabular}
\label{tab:ttest_precision}
\end{table}

\begin{table}[ht]
\centering
\caption{T-Test Results: Recall (QSVM-QNN vs Others)}
\begin{tabular}{lcc}
\toprule
\textbf{Compared With} & \textbf{t-statistic} & \textbf{p-value} \\
\midrule
KNN & 22.66 & 1.11e-14 \\
RF & 20.04 & 9.28e-14 \\
SVM & -6.39 & 5.08e-06 \\
CNN & 2.35 & 3.02e-02 \\
LR & -10.04 & 8.34e-09 \\
DT & 77.43 & 3.58e-24 \\
NB & 99.40 & 4.04e-26 \\
QNN & 1672.86 & 3.49e-48 \\
QSVM & 35.66 & 3.75e-18 \\
\bottomrule
\end{tabular}
\label{tab:ttest_recall}
\end{table}

\begin{table}[ht]
\centering
\caption{T-Test Results: F1 (QSVM-QNN vs Others)}
\begin{tabular}{lcc}
\toprule
\textbf{Compared With} & \textbf{t-statistic} & \textbf{p-value} \\
\midrule
KNN & -30.56 & 5.78e-17 \\
RF & -47.91 & 1.94e-20 \\
SVM & -88.78 & 3.07e-25 \\
CNN & -55.54 & 1.39e-21 \\
LR & -67.30 & 4.43e-23 \\
DT & 25.92 & 1.05e-15 \\
NB & 313.59 & 4.27e-35 \\
QNN & 1926.85 & 2.74e-49 \\
QSVM & -52.38 & 3.94e-21 \\
\bottomrule
\end{tabular}
\label{tab:ttest_f1}
\end{table}

\begin{table}[ht]
\centering
\caption{Accuracy Comparison of Algorithms for EEG Datasets}
\begin{tabular}{l l c}
\toprule
\textbf{Algorithm} & \textbf{Dataset} & \textbf{Accuracy} \\
\midrule
SVM~\cite{duan2013differential}         & SEED        & 0.867 \\
KNN~\cite{duan2013differential}         & SEED        & 0.706 \\
KNN~\cite{kastrati2021eegeyenet}        & EEGEyeNet   & 0.907 \\
CNN~\cite{kastrati2021eegeyenet}        & EEGEyeNet   & 0.908 \\
XGBoost~\cite{kastrati2021eegeyenet}    & EEGEyeNet   & 0.979 \\
EEG modality~\cite{koelstra2012deap}    & DEAP        & 0.620 \\
FBCNet~\cite{ma2022large}               & EEG-BIDS    & 0.688 \\
QSVM                                     & EEGMM       & 0.987 \\
QSVM                                     & EEG         & 0.998 \\
QSVM-QNN                                 & EEGMM       & 0.990 \\
QSVM-QNN                                 & EEG         & 0.950 \\
\bottomrule
\end{tabular}
\label{table:accuracy_comparison}
\end{table}

\subsection{Noisy Results}
\begin{figure}
    \centering
    \begin{subfigure}{0.48\textwidth}
        \includegraphics[width=\linewidth]{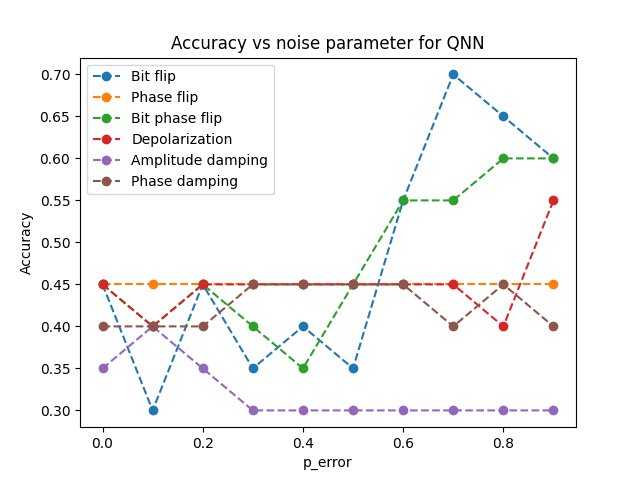}
        \caption{}
        \label{fig:sub1}
    \end{subfigure}\hfill
    \begin{subfigure}{0.48\textwidth}
        \includegraphics[width=\linewidth]{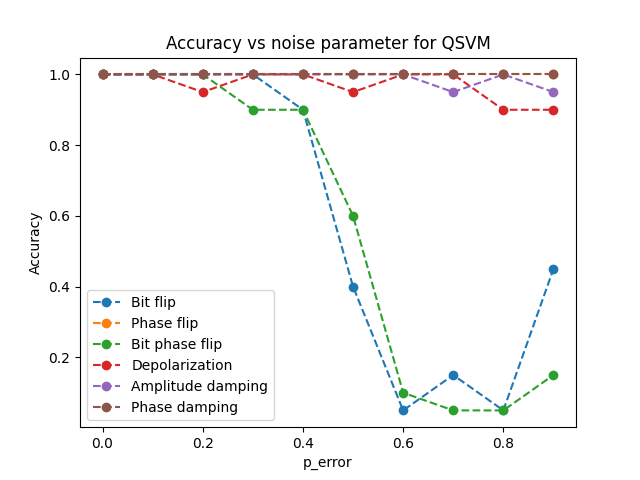}
        \caption{}
        \label{fig:sub2}
    \end{subfigure}
    \begin{subfigure}{0.48\textwidth}
        \includegraphics[width=\linewidth]{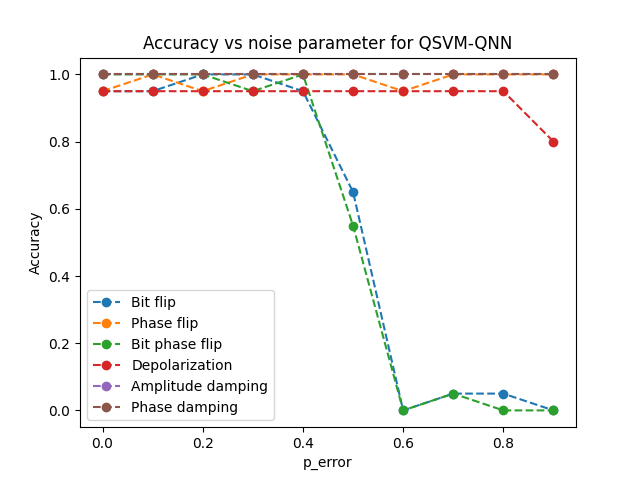}
        \caption{}
        \label{fig:sub1}
    \end{subfigure}\hfill
    \caption{Accuracy Comparison across Different Noise Models for the EEGMM Dataset: (a) QNN, (b) QSVM, and (c) QSVM-QNN.}
    \label{fig6:}
\end{figure}

The performance of QNN, QSVM, and QSVM-QNN against noise models in the EEGMM dataset is shown in Fig. \ref{fig6:}. The PF model does not affect accuracy, making it an ideal option for QNN. Whereas AD produces the least accuracy as noise levels increase, DP and PD vary slightly as noise parameters rise.
It is interesting to note that both BF and BPF improve accuracy as noise parameter values increase. QSVM and QSVM-QNN exhibit similar accuracy variations in varying noise factors. QSVM, PF, and PD maintain consistent accuracies as the noise parameters grow; however, AD and DP fluctuate slightly. After the noise parameters reach 0.4, the BF and BPF accuracies begin to degrade, and by 0.6, the accuracies are almost zero. They tend to rise after the noise parameter value reaches 0.8. However, the QSVM-QNN exhibits a similar but slightly distinct pattern of accuracies over the noise parameters. PD and AD noise have constant accuracies across all noise parameters, whereas PF and DP accuracies fluctuate slightly. After a noise value of 0.4, the BF and BPF accuracies decrease and remain close to zero at 0.6.

\begin{figure}
    \centering
    \begin{subfigure}{0.48\textwidth}
        \includegraphics[width=\linewidth]{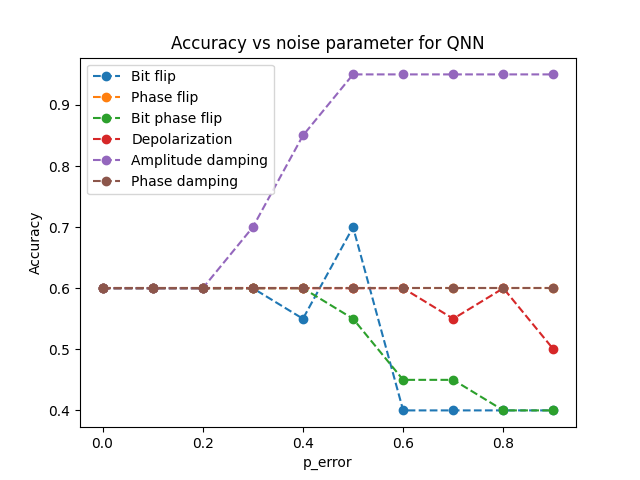}
        \caption{}
        \label{fig:sub1}
    \end{subfigure}\hfill
    \begin{subfigure}{0.48\textwidth}
        \includegraphics[width=\linewidth]{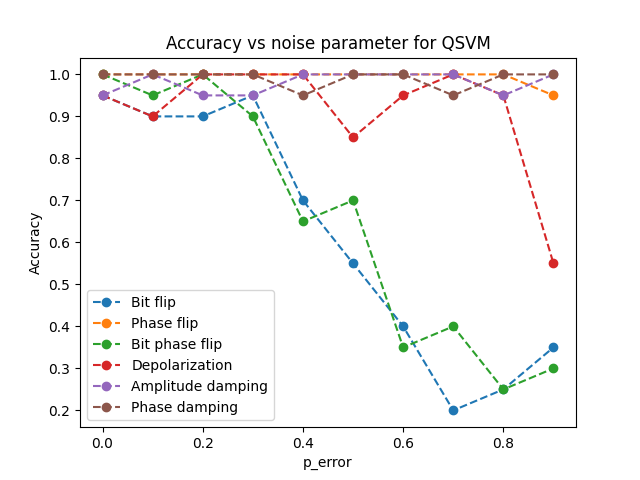}
        \caption{}
        \label{fig:sub2}
    \end{subfigure}
    \begin{subfigure}{0.48\textwidth}
        \includegraphics[width=\linewidth]{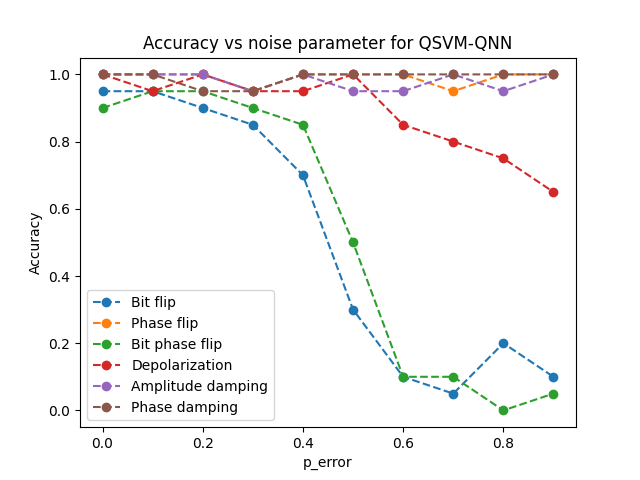}
        \caption{}
        \label{fig:sub1}
    \end{subfigure}\hfill
    \caption{Accuracy Comparison across Different Noise Models for the EEG Dataset: (a) QNN, (b) QSVM, and (c) QSVM-QNN}
    \label{fig7:}
\end{figure}

Fig. \ref{fig7:} shows the accuracy of the QNN, QSVM, and QSVM-QNN algorithms on the EEG dataset. In the case of QNN, PD and PF maintain constant accuracies as noise parameters increase, whereas DP produces a small deviation after a noise parameter of 0.6. After the noise parameter of 0.4, the accuracy of BF and BPF decreases, and after the noise parameter of 0.6, the accuracy remains around 0.4. AD increases accuracy after noise parameter 0.2, then crosses the accuracy threshold of 0.9 and remains constant as the noise parameter grows. It can be mentioned that, while QNN has the lowest accuracy of roughly 0.6 for the EEG dataset, in the presence of AD, it achieves an accuracy of more than 0.9, making it an excellent candidate for this algorithm. For QSVM, PD has no effect on accuracy, whereas PF, AD, and DP cause accuracy to fluctuate. The poorest models are BF and BPF, which exhibit a drop in accuracy when the noise parameter value increases after 0.4, which brings them close to zero. QSVM-QNN behaves similarly; PD, AD, PF, and DP exhibit minor accuracy fluctuations; however, BF and BPF approach zero when the noise parameter exceeds 0.6. In this scenario, PD has the least volatility, making it an excellent choice for this algorithm.

\subsection{Discussion}
Here, we summarize several inferences that are observed from the above results while performing the experiments on the two EEG datasets. There are no significant changes in accuracy while increasing or reducing the number of dimensions in the two datasets. With a minimum number of features, both datasets have better accuracy in comparison to the existing algorithms. Taking more QNN layers does not change or enhance the accuracy significantly, so only 5 layers of QNN have been taken in the cases of QNN and QSVM-QNN. QSVM and QSVM-QNN give similar accuracies, but the circuit depth for QSVM-QNN is larger due to the addition of a parametrized circuit. Both QNN and QSVM take less time to execute, while QSVM-QNN takes longer due to its large circuit depth. To provide practical experimental support for the claim of reduced computational cost, the execution time was benchmarked for all models. The QNN model completed training in 1.2 minutes, QSVM in 1 minute, while QSVM-QNN took around 2 minutes for the same dataset, confirming the added overhead due to the depth of the circuit. These results support the claim that QNN and QSVM incur lower computational costs compared to deeper quantum models.

Most of the classical algorithms, QSVM and QSVM-QNN, give 1.000 accuracies for 100 data points. Both QSVM and QSVM-QNN exhibit almost similar behavior in terms of accuracy when running them in the presence of noise. For QNN, BF and BPF noises help increase the accuracy for the EEGMM dataset, whereas AD noise helps increase the accuracy for the EEG dataset. For QSVM and QSVM-QNN, only the BF and BPF models degrade in accuracy, while other noise models either remain constant or fluctuate a little, making both algorithms suitable. PD is the most robust model for QNN, QSVM, and QSVM-QNN algorithms. 

The findings of this study offer valuable insight into the performance of quantum and classical algorithms on EEG datasets, providing a foundation for advancements in BCI systems. Although increasing or reducing the number of dimensions in the datasets did not significantly impact accuracy, the superior performance of algorithms with minimal features highlights their efficiency in handling high-dimensional data, a crucial aspect in BCI systems. The observation that QNN and QSVM-QNN maintain high accuracy even in noisy environments reinforces their potential for robust real-world applications, particularly in settings where EEG signals are often contaminated with noise. Furthermore, the ability of specific noise models, such as BF, BPF, and AD, to influence accuracy differently between datasets underscores the importance of adapting noise handling strategies to specific applications.

The reduced computational overhead achieved by employing a single-layer QNN structure with one parametrized circuit part in the proposed method, compared to multi-layer configurations with multiple parametrized circuit parts ~\cite{bib_Pablo}, demonstrates a significant step forward in making quantum-enhanced BCI systems more feasible for practical deployment. While the QSVM-QNN framework demonstrates improved accuracy and robustness over classical and standalone quantum models, it does have certain limitations. First, the increased circuit depth introduced by combining QSVM with QNN leads to higher quantum resource requirements and longer training times, especially when scaled to larger datasets or deployed on near-term quantum hardware with limited coherence times. Second, the model has been tested using quantum simulators; therefore, its resilience to real hardware-induced noise and decoherence remains to be fully validated.
\section{Conclusion}\label{Sec5}
The advancement of BCI systems offers significant potential to transform human-computer interaction and assist individuals with disabilities. However, these systems face significant challenges, including signal variability, limited accuracy, and the need for user training. Although current algorithms have addressed some of these issues, they often struggle to handle complex signal patterns and achieve real-time performance. 
In this study, three QML algorithms, QNN, QSVM, and the proposed QSVM-QNN, were implemented on the two EEG-based BCI datasets. Their performance was evaluated and compared against the classical algorithms, including KNN, RF, SVM, CNN, LR, DT, and NB, using various metrics. The results show that QSVM and QSVM-QNN achieved a comparable or superior accuracy to the classical methods. The QSVM-QNN achieved the highest accuracy of 0.990 in the EEGMM dataset, while the QSVM performed best in the EEG dataset.
Furthermore, the robustness of the quantum algorithms against various noise models is verified. PD is found to be the most robust noise model, whereas the BF and BPF models significantly affected the accuracy of QSVM and QSVM-QNN. Other models showed minor fluctuations, suggesting the quantum algorithms' general robustness under noisy conditions.
In future work, the proposed quantum algorithms will be tested on real quantum hardware platforms such as IBM Q or IonQ to evaluate their practical performance in realistic noise environments. This will help validate simulation results and identify implementation challenges, enabling hardware-specific optimizations.

\normalsize


\ifCLASSOPTIONcaptionsoff
  \newpage
\fi



\bibliographystyle{IEEEtran}
%

\bibliography{reference}

\vfill


\end{document}